\documentclass[prd,aps,showpacs,nofootinbib,showkeywords,elsarticle, eqsecnum,superscriptaddress]{revtex4-2}
\usepackage{graphicx,color,amsmath,amsxtra}
\usepackage{epsf}
\usepackage{amssymb}
\usepackage{enumerate}
\usepackage{hhline}
\usepackage{array}
\usepackage{tabularx}
\usepackage[colorlinks=true, citecolor=blue, urlcolor = blue, linkcolor= red, bookmarks=true]{hyperref}
\usepackage{graphicx}                
\usepackage{epstopdf}
\usepackage{orcidlink}
\begin{document}
\pagestyle{myheadings}

\title{Thermodynamics and phase transitions of charged-AdS black holes in dRGT massive gravity with nonlinear electrodynamics.}
\author{Mohd Rehan \texorpdfstring{\href{https://orcid.org/0009-0003-1552-7817}{\orcidlink{0009-0003-1552-7817}}{}}}
\email{rk1958227@gmail.com}
\affiliation{Centre for Theoretical Physics, Jamia Millia Islamia, New Delhi 110025, India}
\author{Arun Kumar \texorpdfstring{\href{https://orcid.org/0000-0001-8461-5368}{\orcidlink{0000-0001-8461-5368}}{}}} \email{arunbidhan@gmail.com} 
\affiliation{Institute for Theoretical Physics and Cosmology, Zhejiang University of Technology, Hangzhou 310023, China}
\author{Tuan Q. Do \texorpdfstring{\href{https://orcid.org/0000-0002-4559-5162}{\orcidlink{0000-0002-4559-5162}}{}}}
\email{tuan.doquoc@phenikaa-uni.edu.vn}
\affiliation{Phenikaa Institute for Advanced Study, Phenikaa University, Hanoi 12116, Vietnam}
\author{Sushant~G.~Ghosh \texorpdfstring{\href{https://orcid.org/0000-0002-0835-3690}{\orcidlink{0000-0002-0835-3690}}{}}}\email{sghosh2@jmi.ac.in}
\affiliation{Centre for Theoretical Physics, 
Jamia Millia Islamia, New Delhi 110025, India}
\affiliation{Astrophysics and Cosmology Research Unit, 
School of Mathematics, Statistics and Computer Science, University of KwaZulu-Natal, Private Bag 54001, Durban 4000, South Africa}
\date{\today} 

\begin{abstract}
Investigating black holes in modified theories of gravity offers fertile ground for exploring phenomena beyond the scope of general relativity. We investigate a novel class of charged anti-de Sitter (AdS) black holes within the ghost-free de Rham–Gabadadze–Tolley (dRGT) massive gravity, minimally coupled to an exponential form of nonlinear electrodynamics (NED).  The NED sector is modelled by an exponential electrodynamics Lagrangian, which leads to singular black hole geometries in contrast to many regular configurations known in other NED models. In turn, we systematically investigate the thermodynamic properties and phase structure of the obtained black holes. The results show that the system has a rich thermodynamic structure. For different values of the magnetic charge $q$, the black hole can exhibit several types of phase transitions. These include van der Waals–like first-order phase transitions, second-order critical behavior, and a reentrant phase transition between small and large black holes without extending the phase space ($\Lambda=$constant).  Our study enhance the understanding of AdS black holes in ghost-free massive gravity, providing further insights into the interplay between graviton mass and NED. The results highlight how the combined effects of graviton mass and electromagnetic nonlinearity can yield a rich and complex thermodynamic phase space, offering further insights relevant to the gauge/gravity duality and the ongoing search for observational signatures of modified gravity.\\

\noindent\textbf{Keywords:} dRGT massive gravity, nonlinear electrodynamics, charged AdS black holes, thermodynamics, phase transitions, reentrant phase transitions, and van der Waals behaviour. 
\end{abstract}

\pacs{04.50.Kd, 04.70.-s,  04.70.Dy, 04.20.Jb }

\maketitle

%%%%%%%%%%%%%%%%%%%%%%%%%%%%%%%%%%%%%%%%%%%%%%%%%%%%%%
\section{Introduction} \label{intro}
Massive gravity has a long and rich history, dating back to the seminal paper by Fierz and Pauli (FP) \cite{Fierz:1939ix}. In fact, this theory had not been widely considered in the cosmology community for several decades. The main reason is due to the vDVZ discontinuity \cite{vanDam:1970vg,Zakharov:1970cc}, i.e., it will not recover Einstein's general relativity in the massless limit as pointed out by van Dam and Veltman as well as by Zakharov in Refs. \cite{vanDam:1970vg,Zakharov:1970cc}. In principle, this vDVZ discontinuity can be resolved in nonlinear extensions of the FP theory as claimed by Vainshtein in Ref.  \cite{Vainshtein:1972sx}. Unfortunately, these nonlinear levels introduce a ghost, which has negative kinetic energy and renders the corresponding theory unstable, as shown by Boulware and Deser \cite{Boulware:1972yco}. As a result,  the emergence of the BD ghost is due to the existence of an extra mode, i.e., the sixth mode of the massive graviton, which cannot be eliminated due to the lack of a Hamiltonian constraint in the Arnowitt-Deser-Misner (ADM) language \cite{Arnowitt:1962hi}. Hence, building a nonlinear but ghost-free massive gravity has been a great challenge to physicists for several decades. In fact, there have been significant efforts to develop such a theory, but none of them have been entirely successful \cite{Arkani-Hamed:2002bjr,Dubovsky:2004sg,Creminelli:2005qk,Hinterbichler:2011tt}. An interesting review paper on this theory can be seen in Ref. \cite{Hinterbichler:2011tt}.

Recently, the \textit{de Rham--Gabadadze--Tolley} (dRGT) model has successfully constructed a nonlinear but ghost-free massive theory \cite{deRham:2010ik,deRham:2010kj}. Several different proofs for this ghost-free feature of the dRGT theory, even when the reference metric is arbitrary, have been demonstrated \cite{Hassan:2011hr,Hassan:2011ea,Hassan:2011tf}. Very soon after the seminal papers of dRGT group,  a number of black hole solutions have been found for the dRGT massive gravity \cite{Nieuwenhuizen:2011sq,Koyama:2011xz,Berezhiani:2011mt,Cai:2012db,Babichev:2014fka}. In his interesting paper \cite{Vegh:2013sk}, Vegh has shown that the dRGT massive gravity can act as a holographic framework for translational symmetry breaking
and momentum dissipation. As a result, he arrived at a holographic theory of solids based
on Lorentz-breaking graviton mass terms. As a consequence, this paper initiated a number of follow-up studies on the so-called holographic massive gravity \cite{Blake:2013bqa,Blake:2013owa,Davison:2013jba,Adams:2014vza}. It has inspired many other studies on charged AdS black holes within the context of dRGT massive gravity \cite{Cai:2014znn,Zhou:2015dha,Xu:2015rfa,Zou:2016sab,Hendi:2015pda,Hendi:2015bna,Hendi:2016yof,Dehyadegari:2017fqo,Dehghani:2019thq,Dehghani:2020blz,Ghosh:2015cva,Tannukij:2017jtn,Nam:2018ltb,Ghosh:2019eoo,Paul:2023mlh,Brito:2013xaa,Tolley:2015ywa,Ditta:2025ksr}.  It is worth noting that many other interesting works on black hole physics of the dRGT theory can be seen in an interesting review by de Rham \cite{deRham:2014zqa}. 

It can be stated that the dRGT massive gravity is a typical example of the importance of nonlinearity in physics, as well as in cosmology.  Another example that can be mentioned is the so-called NED, which stems from
 the seminal paper of Born and Infeld published in 1934 \cite{Born:1934gh}. It appears that many nonlinear electromagnetic field models have been proposed and studied extensively over the last three decades  \cite{Soleng:1995kn,AyonBeato:1998ub,Ayon-Beato:1999kuh,Bronnikov:2000vy,AyonBeato:2000zs,Hassaine:2008pw,Gonzalez:2009nn,Hendi:2012um,Fernando:2016ksb,Ghosh:2018bxg,Ali:2018boy,Kumar:2018vsm,Hyun:2019gfz,Nomura:2020tpc,Ghosh:2021clx,Ghosh:2020ijh,Hendi:2020knv,Guo:2021zxl,Rehan:2024dsg,Vachher:2024ldc,Kumar:2020xvu}. Interestingly, a number of novel charged black holes with rich thermodynamic properties and implications have been found in the context of NED. One typical example worth mentioning is the so-called regular black holes, which do not have the spatial singularity at $r=0$, have been shown to exist within the context of NED \cite{AyonBeato:1998ub,Ayon-Beato:1999kuh,Bronnikov:2000vy,AyonBeato:2000zs,Fernando:2016ksb,Ghosh:2018bxg,Ali:2018boy,Kumar:2018vsm,Hyun:2019gfz,Nomura:2020tpc,Ghosh:2020ijh,Hendi:2020knv,Guo:2021zxl,Rehan:2024dsg,Kumar:2020bqf,Ghosh:2020tgy,Kumar:2024qon,Kumar:2024sdg,Kumar:2023ijg}.
 
Inspired by these two nonlinear theories, some people have proposed to seek charged black hole solutions for their combinations, in which the dRGT massive gravity is allowed to be minimally coupled to NED \cite{Hendi:2015bna,Hendi:2016yof,Dehyadegari:2017fqo,Ghosh:2019eoo,Nam:2018ltb,Paul:2023mlh}.  In harmony with these research directions, we consider in this paper a model of the dRGT massive gravity minimally coupled to a nonlinear electromagnetic field. As a result, a novel class of exactly charged AdS black holes will be figured out in this proposed model. Very interestingly, the obtained black holes are not regular, in contrast to previous ones found in the context of NED \cite{AyonBeato:1998ub,Ayon-Beato:1999kuh,Bronnikov:2000vy,AyonBeato:2000zs,Fernando:2016ksb,Ghosh:2018bxg,Ali:2018boy,Kumar:2018vsm,Hyun:2019gfz,Nomura:2020tpc,Ghosh:2020ijh,Hendi:2020knv,Guo:2021zxl,Rehan:2024dsg}. This is due to a result that the massive graviton potentials act as an effective cosmological constant $\Lambda$. In addition, it will be shown that the obtained charged AdS black holes turn out to be different from ones derived previously in Refs. \cite{Ghosh:2015cva,Nam:2018ltb} in some limits. Inspired by many interesting works on thermodynamics of charged AdS  black holes within the context of dRGT massive gravity \cite{Cai:2014znn,Zhou:2015dha,Xu:2015rfa,Zou:2016sab,Hendi:2015pda,Hendi:2015bna,Hendi:2016yof,Dehyadegari:2017fqo,Dehghani:2019thq,Dehghani:2020blz,Ghosh:2015cva,Tannukij:2017jtn,Nam:2018ltb,Ghosh:2019eoo,Paul:2023mlh}, we will analyze the thermodynamics of the obtained charged AdS black holes.  It should be noted that the reason for focusing on AdS black holes rather than dS black holes is due to two points: (i) First, it is known that AdS black holes are thermodynamically stable, as proved by Hawking and Page \cite{Hawking:1982dh}. (ii) Second, it has been shown that AdS black holes are relevant for studies of the so-called AdS/CFT correspondence, firstly proposed by Maldacena \cite{Maldacena:1997re,Gubser:1998bc,Witten:1998qj}.

Now, let us take a moment to discuss briefly the second point mentioned above. It appears, according to the holographic dictionary discovered in Refs.~\cite{Maldacena:1997re,Gubser:1998bc,Witten:1998qj}, that a bulk AdS black hole (with gravity) will correspond to a boundary finite temperature conformal field theory (without gravity). This interesting point suggests an important consequence: a strongly coupled field theory, which can be very difficult to control, can be handled by investigating the corresponding weakly coupled classical gravity theory \cite{Maldacena:1997re,Gubser:1998bc,Witten:1998qj}. There have been two well-known evidences for this claim. First, it showed in Ref. \cite{Hawking:1982dh} that the so-called Hawking-Page phase transition, in which AdS black holes undergo to a thermal AdS space under a certain Hawking temperature, can be identified with the confinement or deconfinement phase transition in large-$N$ gauge theory as pointed out by Witten in Ref. \cite{Witten:1998zw}. When AdS black holes become charged, a first-order phase transition between large black holes and small black holes will exist, provided that the charge is smaller than a critical value. Amazingly, this phase transition of charged AdS black holes is very similar to a Van der Waals liquid-gas phase transition as demonstrated by Chamblin et al. in Refs. \cite{Chamblin:1999tk,Chamblin:1999hg}.

Despite significant progress in studying black hole solutions in massive gravity and NED separately, their combined effects remain comparatively not much explored. Exponential NED, in particular, offers a physically motivated framework for introducing nonlinear corrections to the standard Maxwell field.  Motivated by these ideas, we study charged black hole solutions in the context of dRGT enormous gravity minimally coupled to exponential NED in this paper. Our main goal is to understand how the spacetime geometry and the effective gravitational potential resulting from massive gravity are altered by the NED. In particular, we obtain an excat black hole solution and systematically examine how the exponential nonlinear term shapes the metric structure and the thermodynamic behavior.

The present paper is organised as follows: The motivations for our current study are presented in Section \ref{intro}.  Section \ref{sec2} establishes our theoretical framework, presenting the action for dRGT massive gravity coupled to NED and deriving the corresponding field equations. We then provide a comprehensive thermodynamic analysis. We begin by computing fundamental quantities (mass, temperature, entropy), then examine local stability through specific heat in  Subsection \ref{3A}, and finally investigate global stability and phase structure via Gibbs free energy in Subsection \ref{3B}, identifying various phase transitions, including reentrant behaviour, in Section \ref{sec3}. Finally, we summarise our main results, discuss their physical implications, and suggest directions for future research in Section \ref{final}

Throughout this work, we use natural units where $c = \hbar = G = 1$.

%%%%%%%%%%%%%%%%%%%%%%%%%%%%%
%%%%%%%%%%%%%%%
\section{NED charged black holes in dRGT massive gravity} \label{sec2}
Let us consider the dRGT massive gravity minimally coupled to an NED. The dRGT massive gravity is a generalisation of Einstein’s general relativity that successfully avoids the Boulware--Deser ghost through the introduction of carefully constructed graviton potential terms in the action~\cite{deRham:2010kj,deRham:2010ik,Hassan:2011hr}. 
It can be interpreted as Einstein gravity interacting with a nondynamical reference (fiducial) metric, which effectively endows the graviton with a finite mass. 
To study electromagnetic effects beyond the linear Maxwell theory, we introduce a minimal coupling between the dRGT massive gravity and a NED, described by a general Lagrangian $\mathcal{L}(F)$. 
Such NED models, originally proposed by Born and Infeld to remove the divergence of the self-energy of point charges~\cite{Born:1934gh}, have been widely employed to construct regular or nonregular black hole solutions and to explore their rich thermodynamic behavior~\cite{AyonBeato:1998ub,Hendi:2012um,Hendi:2015hoa}. 
The combined framework of dRGT massive gravity coupled to NED, therefore, provides a natural setting to investigate how both the graviton mass and electromagnetic nonlinearity affect the geometry and thermodynamics of charged AdS black holes.  As a result, the general action for the model of dRGT massive gravity minimally coupled to an NED field can be written as
\begin{eqnarray} 
S &=&\int {d^4 x \sqrt {-g}} \Bigl\{  \frac{1}{16\pi} \left[R +m_g^2\left({{\cal L}_2+\alpha_3 {\cal L}_3 +\alpha_4 {\cal L}_4 } \right) \right] -\frac{1}{4\pi}{\cal L}({\cal F}) \Bigr\},
\end{eqnarray}
where $m_g$ is the mass of the graviton, $R$ is Ricci scalar,  $\alpha_{3}$ and $\alpha_4$ are just dimensionless free parameters of the theory that represent the nonlinear interaction terms responsible for maintaining the ghost-free nature of the theory. The quantity ${\cal L}({\cal F})$ is an arbitrary function of ${\cal F} \equiv \frac{1}{4}F_{\mu\nu}F^{\mu\nu}$ with $F_{\mu\nu}\equiv \partial_\mu A_\nu -\partial_\nu A_\mu$ being the field strength of the electromagnetic field. Unlike the standard Maxwell Lagrangian, the NED generalisation allows for richer physical behaviour, such as finite self-energy of point charges and regular or nonregular black hole geometries~\cite{Born:1934gh,AyonBeato:1998ub,Hendi:2012um,Hendi:2015hoa}.  While ${\cal L}_i$ with $i=2-4$ are called massive graviton potentials given by \cite{deRham:2010ik,deRham:2010kj}
\begin{eqnarray}
{\cal L}_2 &=& [{\cal K}]^2- [{\cal K}^2], \\
{\cal L}_3 &=&    \frac{1}{3}  [{\cal K}]^3-[{\cal K}] [{\cal K}^2]+\frac{2}{3}[{\cal K}^3] , \\
{\cal L}_4 &= &  \frac{1}{12}[{\cal K}]^4-\frac{1}{2}[{\cal K}]^2 [{\cal K}^2]+\frac{1}{4} [{\cal K}^2]^2+\frac{2}{3}[{\cal K}][{\cal K}^3]-\frac{1}{2}[{\cal K}^4].
\end{eqnarray}
Here, our model does not involve a pure (negative) cosmological constant, in contrast to many previous studies \cite{Vegh:2013sk,Cai:2014znn,Hendi:2015bna,Nam:2018ltb}. Instead, we will show later that an effective cosmological constant will emerge from the obtained black hole solutions due to the existence of the massive graviton potentials ${\cal L}_i$. This result is indeed consistent with our previous investigation \cite{Ghosh:2015cva}. It should be mentioned that the constant-like behaviour of the massive graviton potentials ${\cal L}_i$ of dRGT massive gravity can be found in many different scenarios, not only in black hole physics \cite{Nieuwenhuizen:2011sq,Koyama:2011xz,Berezhiani:2011mt,Cai:2012db,Babichev:2014fka}  but also in cosmology \cite{deRham:2014zqa, Hinterbichler:2011tt}. 

It is worth noting that these potentials can be constructed using the so-called Cayley-Hamilton (CH) theorem in linear algebra for the determinant of a square matrix \cite{Do:2016abo}. In addition,  we have shown in Ref. \cite{Do:2016abo} using the CH theorem that all massive graviton potentials ${\cal L}_{n+i}$ with $i=1,2,3,...$ must disappear in the $n$-dimensional spacetime. 
In the above expressions, square brackets should be understood as follows~\cite{deRham:2010ik,deRham:2010kj}
\begin{align} 
 [{\cal K}]^n &\equiv \left({\text{tr}{\cal K}^\mu{ }_\nu}\right)^n,\\
  \quad [{\cal K}^n] & \equiv \text{tr}\left( {\cal K}^\mu{ }_{\alpha_1} {\cal K}^{\alpha_1}{ }_{\alpha_2} {\cal K}^{\alpha_2}{ }_{\alpha_3} \ldots {\cal K}^{\alpha_{n-1}}{ }_{\nu} \right),
\end{align}
where ${\cal K}^\mu{ }_{\nu}$ is defined as follows
\begin{equation}
{\cal K}^\mu{ }_\nu \equiv  \delta^\mu{ }_\nu -\sqrt{g^{\mu\alpha} f_{ab}\partial_\alpha \phi^a \partial_\nu \phi^b }.
\end{equation}
In the above definition, $g_{\mu\nu}$ is the (dynamical) physical metric, $f_{ab}$ is the (non-dynamical) fiducial (a.k.a. reference) metric, and $\phi^a$ ($a=0-3$) are the St\"uckelberg scalar fields introduced to ensure the existence of a manifestly diffeomorphism invariance~\cite{Arkani-Hamed:2002bjr}. Since the fiducial metric has been assumed to be non-dynamical, in contrast to the physical metric, any derivatives of scale factors of the fiducial metric will no longer exist in the massive graviton potentials ${\cal L}_{i}$. 

As a result, the corresponding Einstein field equations turn out to be
\begin{equation} \label{Einstein-4d}
\left({R_{\mu\nu}-\frac{1}{2}Rg_{\mu\nu}}\right)+m_g^2 \left({ { X}_{\mu\nu}+ \alpha_4 {Y}_{\mu\nu}}\right)+2g_{\mu\nu}{\cal L} -2 \frac{\partial {\cal L}}{\partial {\cal F}} F_{\mu\gamma} F_\nu{ }^\gamma =0,
\end{equation}
where 
\begin{align} \label{eqX-4d}
X_{\mu\nu}=&  -\frac{1}{2} \left[ \left(\alpha_3+1\right) {\cal L}_2 +\left(\alpha_3+\alpha_4\right) {\cal L}_3 \right] g_{\mu\nu} + \tilde X_{\mu\nu},\\
\tilde X_{\mu\nu}=&~ {\cal K}_{\mu\nu} -[{\cal K}]g_{\mu\nu} - \left(\alpha_3+1\right) \left\{{{\cal K}_{\mu\nu}^2-[{\cal K}]{\cal K}_{\mu\nu} }\right\} +\left(\alpha_3+\alpha_4\right)  \left\{{{\cal K}_{\mu\nu}^3-[{\cal K}]{\cal K}_{\mu\nu}^2+\frac{{\cal L}_2}{2} {\cal K}_{\mu\nu} }\right\}, \\
\label{eqY-4d}
 Y_{\mu\nu} =& -\frac{{\cal L}_4}{2} g_{\mu\nu} + \tilde Y_{\mu\nu},\\
\tilde Y_{\mu\nu} =&~ \frac{{\cal L}_3}{2} {\cal K}_{\mu\nu}  -\frac{{\cal L}_2}{2}  {\cal K}^2_{\mu\nu} +[{\cal K}]{\cal K}^3_{\mu\nu} -{\cal K}^4_{\mu\nu},
\end{align}
where $R_{\mu\nu}$ is Ricci tensor and  ${\cal K}_{\mu\nu}=g_{\mu\alpha_1}{\cal K}^{\alpha_1}{ }_\nu$ and ${\cal K}_{\mu\nu}^n=g_{\mu\alpha_1}{\cal K}^{\alpha_1}{ }_{\alpha_2}... {\cal K}^{\alpha_n}{ }_\nu$ ($n \geq 2$). Note that we showed in Ref.~\cite{Do:2016abo} that $Y_{\mu\nu}$ vanishes automatically for any 4D fiducial and physical metrics as a consequence of the CH theorem. This is a reason why $Y_{\mu\nu}$ has not been mentioned in other papers on the four-dimensional dRGT theory. Hence, we will not consider $Y_{\mu\nu}$  in the rest of this paper. 

In addition, the corresponding field equations of nonlinear electromagnetic field are given by \cite{AyonBeato:2000zs,Fernando:2016ksb,Ghosh:2020ijh}
\begin{equation} \label{EM-field-equation}
\nabla_\mu \left[ \frac{\partial {\cal L}}{\partial {\cal F}} F^{\mu\nu} \right]=0
\end{equation}
and
\begin{equation}\label{EM-field-equation-2}
\nabla_\mu \left(\ast F^{\mu\nu}\right) =0.
\end{equation}

 In order to seek the 4D black hole solutions, the physical metric will be chosen as \cite{Davison:2013jba,Adams:2014vza,Cai:2014znn,Zhou:2015dha, Xu:2015rfa, Zou:2016sab,Hendi:2015pda,Hendi:2015bna,Hendi:2016yof,Dehyadegari:2017fqo, Dehghani:2019thq,Ghosh:2015cva}
 \begin{equation} \label{4D-physical-metric}
 g_{\mu\nu} =  {\text {diag}} \left\{-N(r),F^{-1}(r),r^2,r^2 \sin^2 \theta \right\}
 \end{equation}
 along with the reference metric defined in the spherical coordinates $(t,r,\theta,\varphi)$  as \cite{Davison:2013jba,Adams:2014vza,Cai:2014znn,Zhou:2015dha,Xu:2015rfa,Zou:2016sab,Hendi:2015pda,Hendi:2015bna,Hendi:2016yof,Dehyadegari:2017fqo,Dehghani:2019thq,Ghosh:2015cva}
  \begin{equation} \label{4D-physical}
 f_{ab}= {\text {diag}} \left(0,0,c^2,c^2 \sin^2 \theta\right),
 \end{equation}
where $c$ is an arbitrary constant. Here we assume that $N(r)$ and $F(r)$ are unknown functions of $r$. 

 In order to define the corresponding massive graviton potentials ${\cal L}_i$ ($i=2-4$), we take the unitary gauge, i.e. $\phi^a =x^a$,  for the St\"uckelberg fields, which have been widely chosen in previous papers \cite{Vegh:2013sk,Blake:2013bqa,Blake:2013owa,Davison:2013jba,Adams:2014vza,Cai:2014znn,Zhou:2015dha, Tolley:2015ywa,Xu:2015rfa,Zou:2016sab,Hendi:2015pda,Hendi:2015bna,Hendi:2016yof,Dehyadegari:2017fqo,Dehghani:2019thq,Ghosh:2015cva,Brito:2013xaa}. As a result, we are able to define the non-vanishing components of ${\cal K}^\mu{ }_\nu$ as follows
 \begin{align}
 {\cal K}^0{ }_0 =&~{\cal K}^1{ }_1=1,\\
 {\cal K}^2{ }_2 =&~{\cal K}^3{ }_3= 1-\frac{c}{r}.
 \end{align}
 In addition, the other quantities can be shown to be 
 \begin{align} 
 [{\cal K}]^n =& \left( {\cal K}^0{ }_0 + {\cal K}^1{ }_1 +2 {\cal K}^2{ }_2 \right)^n, \nonumber\\
 [{\cal K}^n] =& \left({\cal K}^0{ }_0 \right)^n +\left({\cal K}^1{ }_1 \right)^n+2 \left({\cal K}^2{ }_2\right)^n .
\end{align}
Given the above results, the massive graviton terms ${\cal L}_{i}$ ($i=2-5$)  are explicitly defined to be
\begin{eqnarray}
{\cal L}_2 &=& 2 \Bigl[ {\cal K}^0{ }_0 \left({\cal K}^1{ }_1 +2 {\cal K}^2{ }_2\right)+ {\cal K}^2{ }_2 \left(2{\cal K}^1{ }_1 +{\cal K}^2{ }_2\right) \Bigr], \\
{\cal L}_3 &=& 2 {\cal K}^2{ }_2 \Bigl[  {\cal K}^0{ }_0 \left(2{\cal K}^1{ }_1 + {\cal K}^2{ }_2\right)+{\cal K}^1{ }_1 {\cal K}^2{ }_2   \Bigr],\\
{\cal L}_4 &=& 2{\cal K}^0{ }_0 {\cal K}^1{ }_1 \left({\cal K}^2{ }_2\right)^2.
\end{eqnarray}
Hence, the corresponding graviton Lagrangian ${\cal L}_M$ turns out to be
\begin{eqnarray} \label{Lagra-reduced}
{\cal L}_M &=&2\left\{ {\cal K}^0{ }_0 {\cal K}^1{ }_1 \left[\alpha_4 \left({\cal K}^2{ }_2\right)^2 +2 \alpha_3 {\cal K}^2{ }_2 +1\right] +{\cal K}^2{ }_2 \left({\cal K}^0{ }_0 + {\cal K}^1{ }_1\right) \left(\alpha_3 {\cal K}^2{ }_2 +2\right) + \left({\cal K}^2{ }_2\right)^2 \right\}. \nonumber\\
\end{eqnarray}
 Thanks to the definition of  $ {\cal K}^\mu{ }_\mu$ given above, the explicit expression of ${\cal L}_M$ can be figured out as
 \begin{align}
 {\cal L}_M = &~\frac{2}{r^2} \left[ {c}^2-6 {c} r+6 r^2 +2 \alpha _3 \left(c-r\right) \left(c-2r\right)  +\alpha _4 \left({c}-r\right)^2 \right].
 \end{align}
 Furthermore, we can expand ${\cal L}_M$ to have a more transparent form,
 \begin{equation}
 {\cal L}_M = 2 \left( \lambda_0 +c \frac{\lambda_1}{r} +c^2\frac{\lambda_2}{r^2} \right),
 \end{equation}
 where
 \begin{align}
 \lambda_0 = &~6+4\alpha_3+\alpha_4, \\
 \lambda_1 =& -2\left(3+3\alpha_3+\alpha_4\right),\\
 \lambda_2 =&~1+2\alpha_3+\alpha_4.
 \end{align}
 Now, we can see that ${\cal L}_M$ will be no longer a constant like $\lambda_0$ but a function of $r$ for $c \neq 0$.  It turns out that ${\cal L}_M$ will be purely constant, i.e. will be equal to $\lambda_0$, once $c$ is set to be zero. This result is consistent with previous investigations in 4D spacetime, e.g., see \cite{Ghosh:2015cva}.  
 Furthermore, the corresponding non-vanishing components of $X_{\mu\nu}$ can be shown to be
 \begin{align}
 X_{00}&=-g_{00}\left(\lambda_0 + c \frac{\lambda_1}{r} +c^2 \frac{\lambda_2}{r^2}\right),\\
 X_{11}&=-g_{11} \left(\lambda_0 + c \frac{\lambda_1}{r} +c^2 \frac{\lambda_2}{r^2}\right),\\
 X_{22}&=- g_{22} \left(\lambda_0 +\frac{c}{2} \frac{\lambda_1}{r} \right), \nonumber\\
 X_{33}&=\sin^2 \theta X_{22}.
 \end{align}
 In addition, the corresponding non-vanishing components of the Einstein tensor, $G_{\mu\nu} \equiv R_{\mu\nu}-\frac{1}{2}g_{\mu\nu} R$, are given by
 \begin{align}
 G_{00} &= g_{00}\frac{1}{r^2} \left(rF'+F-1\right), \\
 G_{11}&= g_{11} \frac{1}{r^2} \left(rF\frac{N'}{N}+F-1\right), \\
 G_{22}&=g_{22}\left[ \frac{F}{2r} \left(r\frac{N''}{N}+\frac{N'}{N} \right) -\frac{N'}{4N} \left(F\frac{N'}{N} -F' \right)+\frac{F'}{2r} \right], \nonumber\\
 G_{33}&=\sin^2\theta G_{22}.
 \end{align}
 In this paper, we propose to consider the exponential electrodynamics \cite{Nam:2018ltb,Ghosh:2018bxg,Ghosh:2020ijh,Kumar:2020cve}
\begin{equation} \label{exponential-electrodynamics}
{\cal L}({\cal F}) = {\cal F} \exp \left[-\frac{k}{q} \left(2q^2 {\cal F} \right)^{1/4} \right],
\end{equation}
 where $k$ is a nonlinear electromagnetic parameter and $q$ is a constant which will be shown later to act as a magnetic charge. It is clear that if $k>0$ along with $q \to 0$ then ${\cal L}({\cal F}) \to 0$ accordingly, meaning that we will no longer have any charged black holes. The same result also happens when $k \to +\infty$, provided that $q>0$; or $k \to -\infty$, provided that $q<0$.  It is noted that one can set $k=0$ at the beginning for obtaining uncharged black holes, but once the corresponding charged black holes are found the limit $k \to 0$ cannot be used to recover uncharged black holes as shown below.  It is also noted that the $D-$dimensional form of  ${\cal L}({\cal F})$ can be found in Ref. \cite{Ghosh:2020ijh}. It is clear that 
 \begin{equation}
 \frac{\partial {\cal L}({\cal F})}{\partial {\cal F}} = \left[1-\frac{k}{4q} \left(2q^2 {\cal F} \right)^{1/4}\right] \exp \left[-\frac{k}{q} \left(2q^2 {\cal F} \right)^{1/4} \right].
 \end{equation}
  In order to figure out black holes, we use the following magnetic ansatz \cite{AyonBeato:2000zs}
  \begin{equation}
  F_{\mu\nu} = 2 \delta^\theta_{[\mu}\delta^\varphi_{\nu]} B(r,\theta),
  \end{equation}
  where $B(r,\theta)$ is unknown function of $r$ and $\theta$. Plugging this ansatz into \eqref{EM-field-equation} we obtain the following solution \cite{AyonBeato:2000zs}
  \begin{equation}
   F_{\mu\nu} = 2 \delta^\theta_{[\mu}\delta^\varphi_{\nu]} q(r) \sin\theta.
   \end{equation}
  Given this solution, \eqref{EM-field-equation-2} leads to the following result \cite{AyonBeato:2000zs}
  \begin{equation}
  q(r)= q = {\text{constant}}.
  \end{equation}
  Note that it has been shown in Ref. \cite{AyonBeato:2000zs} that $q$ appears as a magnetic monopole charge,
  \begin{equation}
  q= \frac{1}{4\pi} \int_{S_2^\infty} {\bf F}, 
  \end{equation}
  with ${\bf F} = \frac{1}{2}F_{\mu\nu} dx^{\mu} \wedge dx^{\nu}$ and $S_2^\infty$ is a sphere at infinity. Hence, we now have all information of the NED field as follows
  \begin{align}
  {\cal F} = &~ \frac{q^2}{2r^4},\\
   {\cal L({\cal F})} =&~ \frac{q^2}{2r^4} \exp \left[ -\frac{k}{r} \right], \\
     \frac{\partial {\cal L}({\cal F})}{\partial {\cal F}}= &~ \left(1-\frac{k}{4r}\right)\exp \left[ -\frac{k}{r} \right].
  \end{align}
  Thanks to these definitions, we are now able to define the corresponding non-vanishing components of Einstein field equation \eqref{Einstein-4d}
  \begin{align}\label{field-equation-1}
  \frac{1}{r^2} \left(rF'+F-1\right) -m_g^2 \left(\lambda_0 + c \frac{\lambda_1}{r} +c^2 \frac{\lambda_2}{r^2}\right) +\frac{q^2}{r^4} \exp \left[ -\frac{k}{r} \right] &=0, \\
  \label{field-equation-2}
  \frac{1}{r^2} \left(rF\frac{N'}{N}+F-1\right) -m_g^2 \left(\lambda_0 + c \frac{\lambda_1}{r} +c^2 \frac{\lambda_2}{r^2}\right) +\frac{q^2}{r^4} \exp \left[ -\frac{k}{r} \right] &=0,\\
  \label{field-equation-3}
\frac{F}{2r} \left(r\frac{N''}{N}+\frac{N'}{N} \right) -\frac{N'}{4N} \left(F\frac{N'}{N} -F' \right)+\frac{F'}{2r}  & \nonumber\\
   -m_g^2 \left(\lambda_0 +\frac{c}{2} \frac{\lambda_1}{r} \right) -\frac{q^2}{r^4}\left(1-\frac{k}{2r}\right)\exp \left[ -\frac{k}{r} \right]&=0.
  \end{align}
  As a result, solving Eq. \eqref{field-equation-1} leads to a non-trivial solution of $F(r)$ as
  \begin{equation} \label{sol-of-F}
  F(r) = 1 + \frac{\mu}{r} -\frac{\Lambda}{3} r^2 + \gamma r +\zeta -\frac{q^2}{kr}\exp \left[ -\frac{k}{r} \right] ,
  \end{equation}
   provided that $k\neq 0$. Here, $\mu$ is an integration constant along with
     \begin{align}
  \label{Lambda}
  \Lambda &= -m_g^2 \lambda_0 = -m_g^2 \left(6+4\alpha_3+\alpha_4 \right), \\
   \label{gamma}
  \gamma &= \frac{1}{2} cm_g^2 \lambda_1 = - cm_g^2 \left(3+3\alpha_3 +\alpha_4 \right), \\
   \label{zeta}
  \zeta &= c^2 m_g^2 \lambda_2= c^2 m_g^2 \left(1+2\alpha_3 +\alpha_4 \right),
  \end{align}
  as new constants defined for convenience, where $\Lambda$ acts as the cosmological constant.  It appears that non-trivial contributions of the massive graviton potentials ${\cal L}_i$ with $i=2-4$ can be clearly observed in terms of the constants $\Lambda$, $\gamma$, and $\zeta$. In addition, it is apparent that the vanishing of $c$, i.e., $c=0$, will imply $\gamma=\zeta =0$; while the vanishing of the mass of the graviton, i.e., $m_g=0$, will lead to $\Lambda=\gamma=\zeta=0$.
 
It turns out that
\begin{equation}
N(r)=F(r),
\end{equation}
after solving Eq. \eqref{field-equation-2} with the help of Eq. \eqref{field-equation-1}. And it is straightforward to confirm that the last equation \eqref{field-equation-3} is satisfied with these solutions. We note that the constant $\Lambda$ depends only on the mass of graviton, $m_g$, and the field parameter $\lambda_0$ as shown in Eq. \eqref{Lambda}. Hence, it can be regarded as an effective cosmological constant, whose value is proportional to the $m_g^2$.  It is well known that $\Lambda <0$, or equivalently $\lambda_0 \equiv (6+\alpha_3+\alpha_4) >0 $, will correspond to  anti-de Sitter (AdS) black holes. On the other hand, $\Lambda>0$, or equivalently $\lambda_0 \equiv (6+\alpha_3+\alpha_4) <0 $, will correspond to de Sitter (dS) black holes.  Note that the integration constant $\mu$ cannot be set to be zero for simplicity. In fact, in connection with the uncharged ($q=0$) Schwarzschild black hole in the dRGT massive gravity found in Ref. \cite{Ghosh:2015cva}, $\mu =-2M$  with $M$ being the mass of the black hole, is a requirement. Hence, we arrive at
    \begin{equation} \label{sol-of-F-new}
  F(r) = 1 - \frac{2M}{r} -\frac{\Lambda}{3} r^2 + \gamma r +\zeta -\frac{q^2}{kr}\exp \left[ -\frac{k}{r} \right].
  \end{equation}
It now becomes clear that non-trivial contributions of the massive graviton potentials ${\cal L}_i$ with $i=2-4$ can be observed in terms of the constants $\Lambda$, $\gamma$, and $\zeta$. 
 
To better understand the role of the NED contribution in Eq.~(\ref{sol-of-F-new}), it is instructive to consider the neutral limit $q=0$. In this case, the nonlinear electromagnetic term vanishes identically, and the metric function reduces to
\begin{equation}
F(r)=1-\frac{2M}{r}-\frac{\Lambda}{3}r^{2}+\gamma r+\zeta ,
\end{equation}
which corresponds to the uncharged black hole solution in dRGT massive gravity \cite{Ghosh:2015cva}. This observation indicates that the exponential term appearing in Eq.~(\ref{sol-of-F-new}) arises solely from the NED and encodes the modification due to the magnetic charge. Therefore, it is expected that the thermodynamic properties of the present charged black hole would be different from those of the black holes found in Ref. \cite{Ghosh:2015cva}. However, in case of $k/r \ll 1$, we have $F(r)$ reduces to 
\begin{equation} \label{approx}
 F(r) \simeq 1 - \frac{2M_{\rm eff}}{r} -\frac{\Lambda}{3} r^2 + \gamma r +\zeta +\frac{q^2}{r^2},
  \end{equation}
  with 
  \begin{equation}
  M_{\rm eff} = M+\frac{q^2}{2k}
  \end{equation}
  acting as an effective mass of the charged black hole. Here, the ratio $q^2/(2k)$ can be regarded as an additional mass contributed to the mass of the black hole.  This result is completely different from the charged black hole found ~\cite{Nam:2018ltb}. The value of $F(r)$ in Eq. \eqref{approx} is distinguishable from that found in Ref. \cite{Ghosh:2015cva} just by the mass of the black hole.  Finally, in the massless graviton limit, i.e., $m_g \to 0$, we have a non-trivial solution,
  \begin{equation}
  F(r) = 1 - \frac{2M}{r} -\frac{q^2}{kr}\exp \left[ -\frac{k}{r} \right].
  \end{equation}
  \begin{figure}[hbtp]
   	\centering
	\includegraphics[scale=0.5]{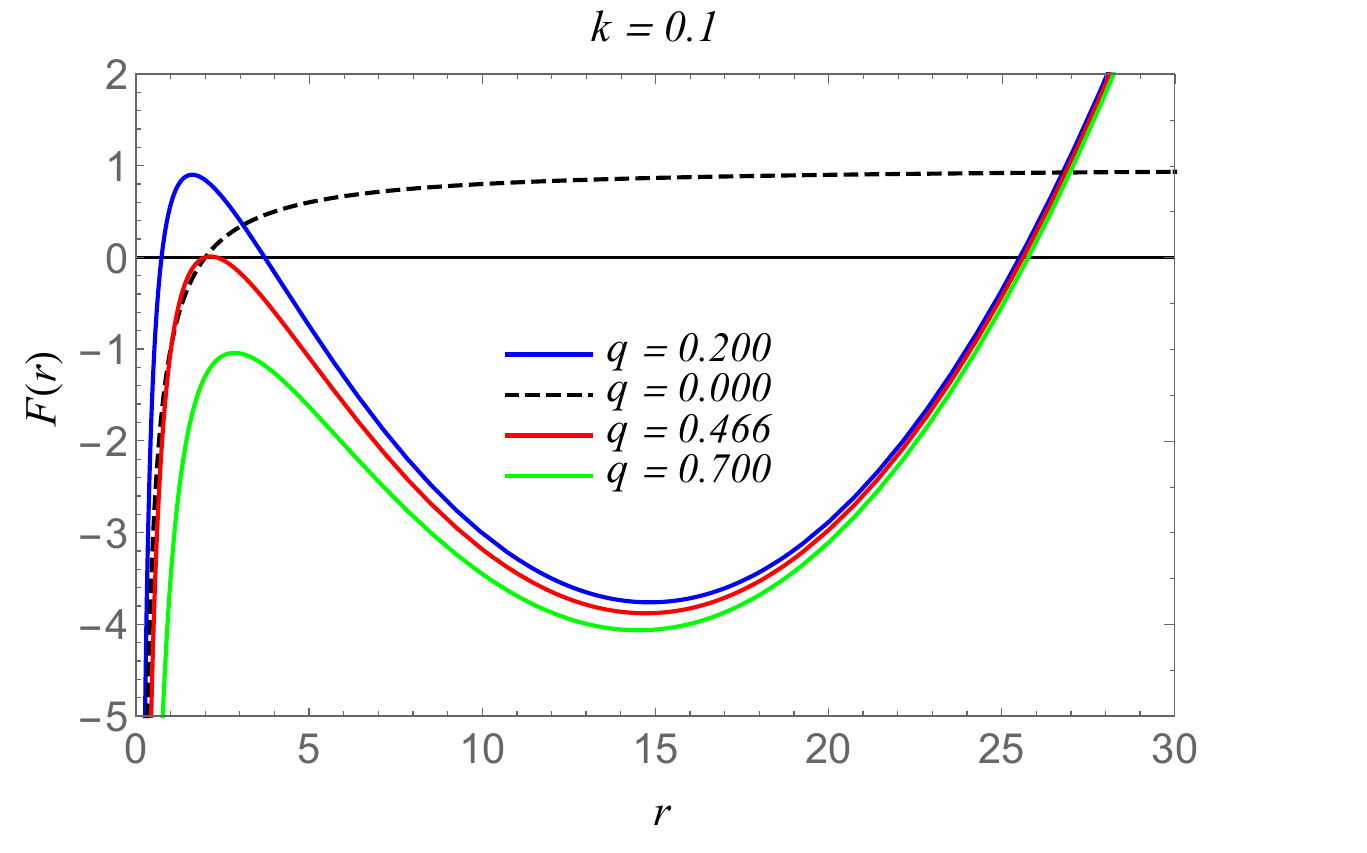}
	\caption{Metric function $F(r)$ vs $r$ for different $q$ with $\gamma = -1$, $\zeta = 2.9$, $k = 0.1$. Curves: $q = 0$ (black dashed), $q = 0.2$ (blue), $q = q_E = 0.466$ (red), $q = 0.7$ (green). The critical charge $q_E$ corresponds to the extremal black hole.} 
	\label{fig1}
\end{figure}
   \begin{figure}[hbtp]
	\centering
	\includegraphics[scale=0.5]{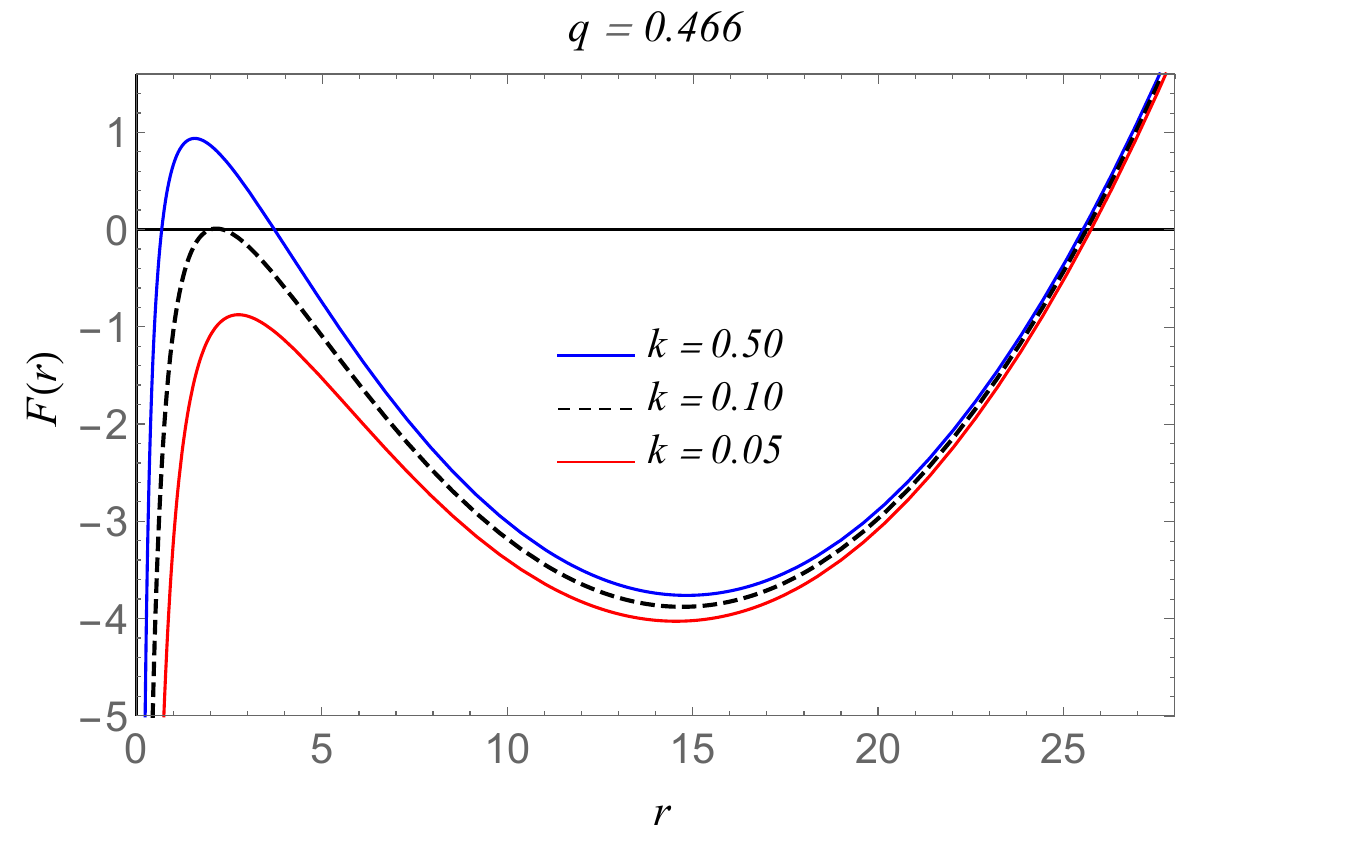}
	\caption{Metric function $F(r)$ vs $r$ for different $k$ with $\gamma = -1$, $\zeta = 2.9$, $q = 0.466$. Curves: $k = 0.05$ (red), $k = 0.1$ (black dashed), $k = 0.5$ (blue). The extremal case corresponds to $k = 0.1$.}
	\label{fig2}
\end{figure}
To end this section, we would like to note that the obtained black holes are not of the regular type, even when the exponential (nonlinear) electrodynamics is present,  in contrast to found before \cite{Ghosh:2018bxg,Ghosh:2020ijh}, since they still contain an unavoidable spatial singularity $r=0$.  The parameters $\gamma$ and $\zeta$, emerging from the massive graviton potential, introduce linear and constant corrections to the metric, respectively, which can be interpreted as modifications to the black hole's effective mass and background spacetime curvature. For concreteness in our numerical analysis, we select the values $\alpha_3 = -3.9$ and $\alpha_4 = 9.7$, which yield $\Lambda<0$ for an AdS spacetime, along with non-trivial $\gamma$ and $\zeta$; we note, however, that the qualitative thermodynamic behaviour, including the van der Waals-like phase transition and reentrant phase transition, is robust across a range of parameter values satisfying the AdS condition.
    %%%%%%%%%%%%%%%%%%%%%%%%%%%%%%%%%%%%%%%%%
  \section{Thermodynamics of charged AdS black hole} \label{sec3}
Having derived the black hole solution in the previous section, we now turn to its thermodynamic analysis. We will focus only on the charged black hole with 
  $\Lambda<0$ due to its rich thermodynamic properties, e.g., see Refs. \cite{Cai:2014znn,Zhou:2015dha, Xu:2015rfa,Zou:2016sab,Hendi:2015pda,Hendi:2015bna,Hendi:2016yof,Dehyadegari:2017fqo,Dehghani:2019thq,Ghosh:2015cva,Ghosh:2020ijh,Paul:2023vys,Ditta:2024tdo,DITTA2024287,Ditta:2024pxl}, especially our recent study \cite{Rehan:2024dsg} for details. 
 First, we must define the corresponding horizon of the found black hole by solving the following equation $F(r_h)=0$, or equivalently,
  \begin{equation} \label{horizon-equation}
   1 - \frac{2M}{r_h} -\frac{\Lambda}{3} r_h^2 + \gamma r_h +\zeta -\frac{q^2}{kr_h}\exp \left[ -\frac{k}{r_h} \right] =0,
   \end{equation}
   here $r_h$ are the horizon radii, which we would like to seek.
  Unfortunately, this equation cannot be solved analytically exactly. Instead, we can only solve it numerically. The same situation can be found in Ref. \cite{Ghosh:2020ijh} for other charged black holes in the presence of exponential electrodynamics. To see quantitatively whether Eq. \eqref{horizon-equation} admits multiple positive roots $r_h$, we are going to plot $F(r)$ as a function of $r$ for several cases of parameters. In particular,  we will choose parameters such as $m_g=1$,  $c=1$, and $M=1$, following Ref. \cite{Ghosh:2015cva}. In addition, the other parameters $\Lambda$, $\gamma$, and $\zeta$ will be determined in terms of $\alpha_3$ and $\alpha_4$ as shown in Eqs. \eqref{Lambda}, \eqref{gamma}, and \eqref{zeta}.  For convenience, we kept $\gamma = -1$, and $\zeta=2.9$ for the horizon structure analysis. We will vary two remaining parameters $q$ and $k$, following Ref. \cite{Ghosh:2020ijh}.  According to Figs. \ref{fig1} and \ref{fig2}, it appears that Eq. \eqref{horizon-equation} can admit three distinct positive roots for a certain set of parameters. Depending on the number of metric function’s root(s), our solution may be a black hole with two
inner and outer horizons, an extreme black hole or a naked singularity.
  
   It is evident from Fig.~\ref{fig1} that the number of horizons decreases as the charge parameter increases. There exists a critical charge $q_E$ corresponding to the extremal black hole with degenerate horizons, for a given set of $m_g$, $c$, $M$, $\Lambda$, $\gamma$, $\zeta$, and $k$. 
   
   Similarly, Fig.~\ref{fig2} corresponds to the given set of $m_g$, $c$, $M$, $\Lambda$, $\gamma$, $\zeta$, and $q$, the number of horizons increases as the nonlinear parameter increases. There exists a critical nonlinear parameter $k_E$ corresponding to the extremal black hole with degenerate horizons. 
 
%%%%%%%%%%%%%%%%%%
\subsection{Basic thermodynamic quantities}\label{3A}
Given the information of the existence of black holes derived in the previous subsection, we are going to investigate the corresponding thermodynamic quantities of the charged black holes. For the thermodynamical analysis of charged black holes in massive gravity, we keep $k=0.1$, $\alpha_3=0.1$, and $\alpha_4=0.2$ throughout. First, the gravitational mass of the found black hole is given by from the equation $F(r_+)=0$ \cite{Ghosh:2020ijh}, 
\begin{equation} \label{def-M-plus}
M_+= \frac{r_+}{2} \left\{  1 -\frac{\Lambda}{3} r_+^2 + \gamma r_+  +\zeta -\frac{q^2}{kr_+}\exp \left[ -\frac{k}{r_+} \right] \right\}.
   \end{equation}
Next, the corresponding temperature of the found black hole can be defined in terms of a relation
\begin{equation}
T =\frac{\kappa}{2\pi},
\end{equation}
where $\kappa$ is nothing but the so-called surface gravity, whose definition is given by
\begin{equation}
\kappa = \left. \left(-\frac{1}{2} \nabla_\mu \xi_\nu \nabla^\mu \xi ^\nu \right)^{1/2} \right|_{r=r_+},
\end{equation}
with $\xi^\mu \equiv \partial/\partial t$ is the timelike Killing vector. Normally, we consider $\xi^\mu =(1,0,0,0)$ along with $\xi_\mu = g_{\mu \nu}\xi^\nu = (-F(r),0,0,0)$ for related calculations. After some simple algebra, we arrive at
\begin{equation}
\kappa =  \left. \frac{1}{2}F' \right|_{r=r_+} = \frac{1}{2} \left [\frac{2M}{r_+^2} +\gamma - \frac{2\Lambda}{3} r_+ + \frac{q^2}{kr_+}\exp \left[ -\frac{k}{r_+} \right] \left( \frac{1}{r_+} -\frac{k}{r_+^2} \right) \right].
\end{equation}
Therefore, the temperature now becomes
\begin{equation} \label{def-T-plus}
T_+ = \frac{1}{4\pi} \left [\frac{2M}{r_+^2} +\gamma - \frac{2\Lambda}{3} r_+ + \frac{q^2}{kr_+}\exp \left[ -\frac{k}{r_+} \right] \left( \frac{1}{r_+} -\frac{k}{r_+^2} \right) \right].
\end{equation}
In the uncharged black hole limit with $q \to 0$, we have
\begin{equation}
T_+ \to T_+^{\rm un} = \frac{1}{4\pi} \left( \frac{2M}{r_+^2} +\gamma - \frac{2\Lambda}{3} r_+ \right).
\end{equation}
Thanks to Eq. \eqref{def-M-plus}, these temperatures are simplified as follows
\begin{align}
T_+ &= \frac{ r_+^2 \left( 1 + 2 r_+ \, \gamma + \zeta - r_+^2 \Lambda \right) - q^2 \exp \left[ -\frac{k}{r_+} \right] }{ 4 \pi r_+^3 }, \\
T_+^{\rm un} &= \frac{1}{ 4 \pi r_+ } \left[ (\zeta + 1) + 2 \gamma r_+ - \Lambda r_+^2 \right].
\end{align}
It appears that $T_+^{\rm un}$ is identical to that defined in Ref.\cite{Ghosh:2015cva,Yasir:2024eyv}. In addition, $T_+$ will recover that of the pure Schwarzschild black hole,
\begin{equation}
T_+ \to T_{\rm Sch} = \frac{1}{4\pi r_+},
\end{equation}
in a massless graviton limit $m_g \to 0$.

\begin{figure}[hbtp]
	\centering
	\includegraphics[width=0.48\textwidth]{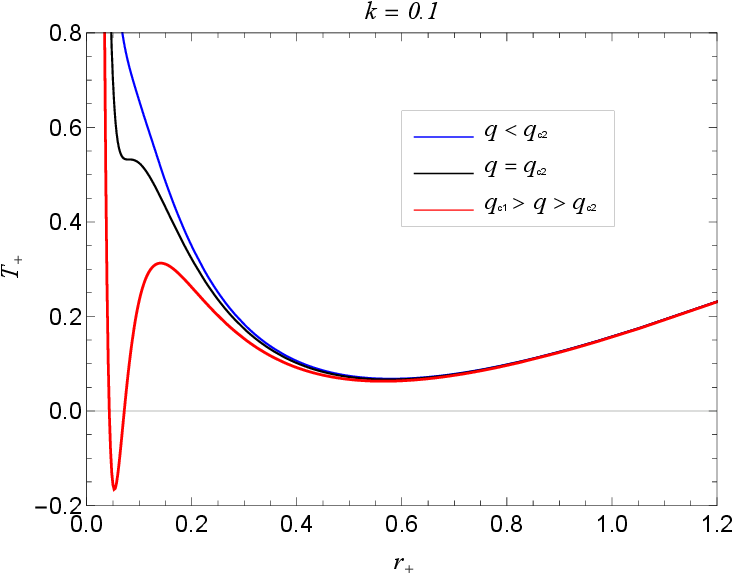}
    \includegraphics[width=0.48\textwidth]{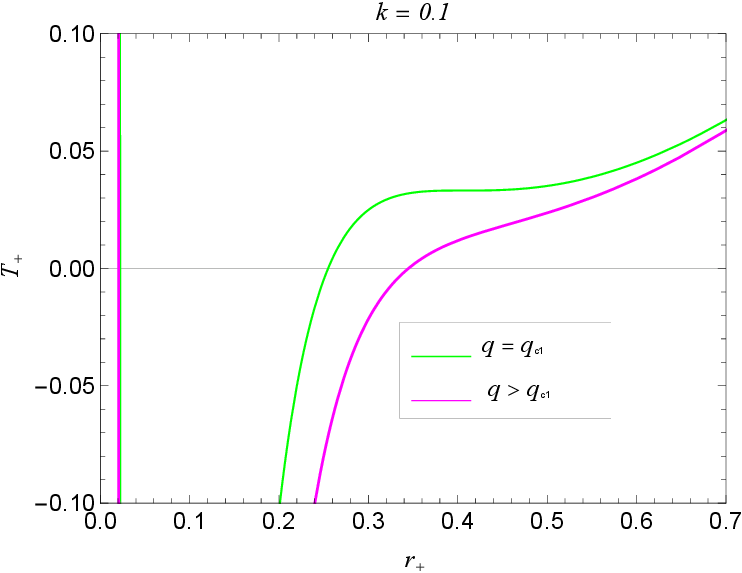}
	\caption{Hawking temperature $T_+$ vs horizon radius $r_+$ for different $q$ with $k = 0.1$, $\alpha_3 = 0.1$, $\alpha_4 = 0.2$. The curves show multiple branches separated by temperature extrema. Critical charges $q_{c1}$ and $q_{c2}$ mark the merging of extrema.}
	\label{fig5}
\end{figure}
\begin{figure}[hbtp]
	\centering
	\includegraphics[width=0.32\textwidth]{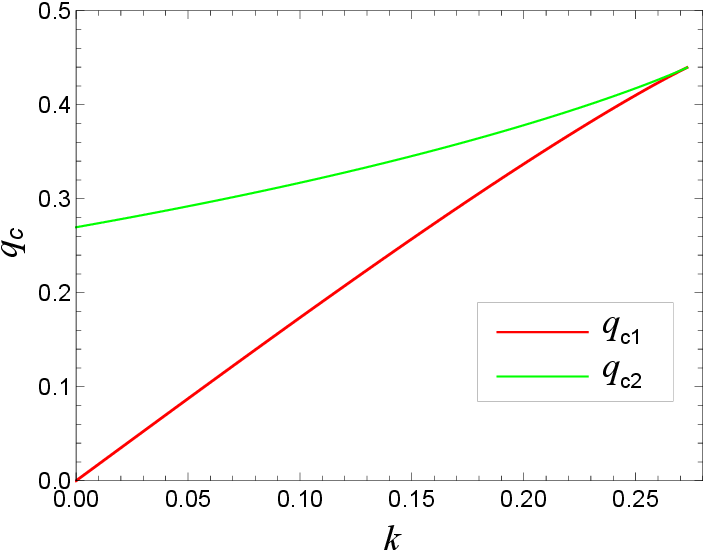}
    \includegraphics[width=0.32\textwidth]{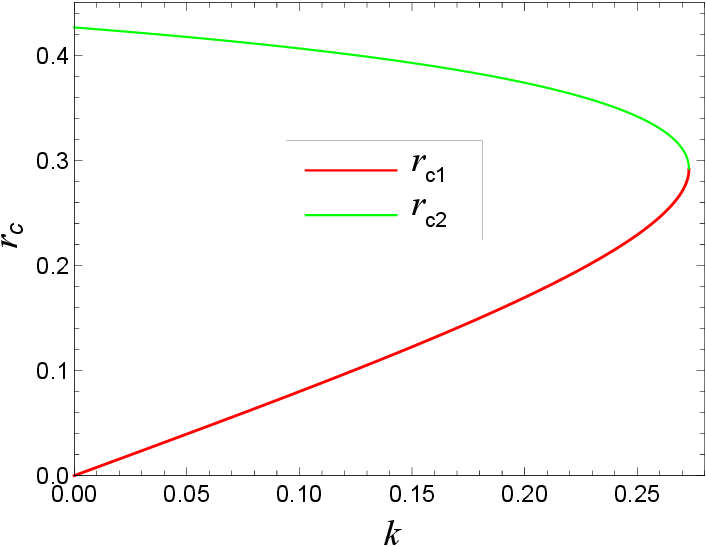}
    \includegraphics[width=0.32\textwidth]{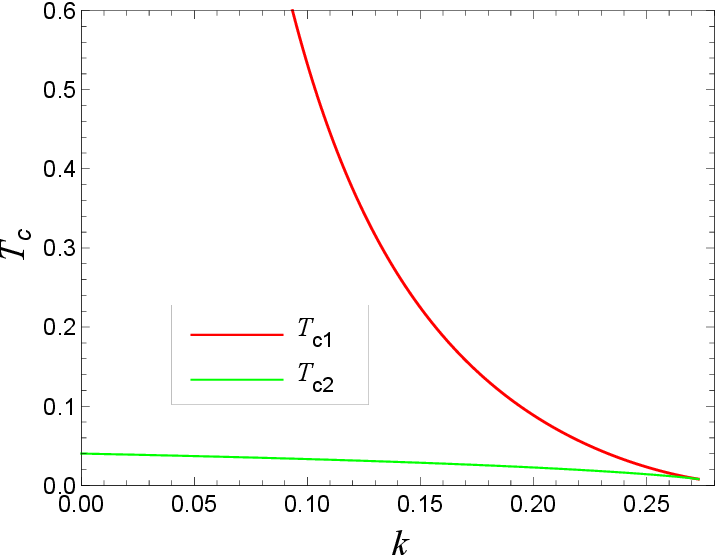}
	\caption{Critical charge $q_c$, horizon radius $r_c$, and temperature $T_c$ vs nonlinear parameter $k$ for $\alpha_3 = 0.1$, $\alpha_4 = 0.2$. Red and green curves correspond to first and second critical points.}
	\label{figcri}
\end{figure}
Incidentally, the limit $k\to 0$ will lead to a result that $T_+ \to T_+^{\rm un}$. However, keep in mind that $k\neq 0$ has been required. It means that the limit $k \to 0$ should not be regarded as a suitable limit for recovering uncharged black holes. We plotted $T_+$ with $r_+$ for different values of charge $q$ in Fig. \ref{fig5}, in which it can be clearly seen that black holes with the same temperature $T_+$ could be described by different horizon radii $r_+$. Depending upon the value of charge $q$, there exist different branches of black holes separated by the extrema of temperature, which can be described as follows:
\begin{itemize}
    \item For $q_{c1}<q<q_{c2}$, there exist three local extrema (two minima and one maximum) which divide the curve into four different branches. Two branches, one ending at the first local minima of the temperature, and the other between maxima and second minima, having a negative slope, are unstable. On the other hand, the other two branches, one starting at the first local minimum and ending at the maximum, and the other starting from the second local minimum, are stable because they exhibit positive slopes. This behaviour can be confirmed from the red curve in Fig. \ref{fig5}.  
    \item For $q=q_{c1}$ and $q=q_{c2}$, the maxima merge with the first and the second maxima, which are represented by the black and green solid curves in Fig. \ref{fig5}, respectively. Hence, there exist three branches for these cases. For $q=q_{c1}$, two branches starting and ending at the merging points are unstable, whereas the branch starting at the second minima is stable. When the charge becomes equal to $q_{c2}$, there exists one unstable branch (ending at the first minimum) and two stable branches ending and starting at the merging point.
    \item For $q<q_{c1}$ and $q>q_{c2}$, there exist only two branches, one stable and one unstable (cf. blue and green curves in Fig. \ref{fig5}). The branch starting at the minima is stable, whereas the branch ending at the minima is unstable.  
\end{itemize} It is also important to mention that for some values of horizon radius, black hole temperature for the branches ending and starting at the first local maximum becomes negative, which is unphysical. Hence, the above-described statements are true only for positive temperature regions. The critical values of charge, $q_{c1}$ and $q_{c2}$, could be found by solving 
\begin{equation}\label{cri}
    \frac{\partial T_+}{\partial r_+}\Bigg|_{q_c,r_c}=0\equiv\frac{\partial^2 T_+}{\partial r^2_+}\Bigg|_{q_c,r_c}.
\end{equation} By using Eq. \eqref{cri}, we can determine the critical horizon radii $r_{c1}$ and $r_{c2}$, and also specify the critical points $(q_{c1},r_{c1},T_{c1})$ and $(q_{c2},r_{c2},T_{c2} )$. For $\alpha_3=0.1$, $\alpha_4=0.2$ and $k=0.1$, the critical points are obtained to be $(0.173482,0.080003,0.532123)$ and $(0.316981,0.406421,0.03321)$. We depicted the behaviour of critical values of charge, horizon radius, and temperature with $k$ in Fig. \ref{figcri} in which the solid red and green curves correspond to the first and second critical points, respectively. The temperature and charge values corresponding to both critical points, respectively, decrease and increase with $k$ before merging with each other at a particular value of $k$. The horizon radii corresponding to the first and second critical points increase and decrease, respectively. It is important to mention that the critical behaviour ceases as we go beyond the value of $k$ where critical points merge.

The next thermodynamic quantity that should be calculated is the corresponding entropy of the charged AdS black holes. We regard these black holes as a canonical ensemble system, where the chemical potential $\phi$ associated with the charge $q$  is held fixed as \cite{Ghosh:2015cva,Ghosh:2020ijh,Feng:2024zxi}
\begin{equation}
 \phi_+ = \frac{q}{r_+}.
 \end{equation}
 Furthermore, these black holes should obey the first law of thermodynamics, stated via the following relation,
 \begin{equation}
 dM_+ =T_+ dS_+ +\phi_+ dq. 
 \end{equation}
 However, since the charge $q$ has been assumed to be constant, then $dq=0$. This means that we simply have
 \begin{equation}
 dM_+ =T_+ dS_+,
 \end{equation}
 which can be integrated out to give the corresponding value of entropy,
 \begin{equation}
 S_+ = \int \frac{dM_+}{T_+},
 \end{equation}
 where $M_+$ and $T_+$ have been defined in Eqs. \eqref{def-M-plus} and \eqref{def-T-plus}, respectively.  
  More specifically, it turns out that
 \begin{align}
 S_+ = 2\pi \int \frac{ 1 - \Lambda r_+^2 +2\gamma r_+ +\zeta - \frac{q^2 }{r_+^2} \exp \left[ -\frac{k}{r_+} \right] }{\frac{2M_+}{r_+^2} +\gamma - \frac{2\Lambda}{3} r_+ + \frac{q^2}{kr_+}\exp \left[ -\frac{k}{r_+} \right] \left( \frac{1}{r_+} -\frac{k}{r_+^2} \right) }dr_+.
 \end{align}
 This integral looks complicated to be done to give an analytic result. Fortunately, using the definition of $M_+$ shown in Eq. \eqref{def-M-plus} can help us to significantly reduce the above expression to a very simple one,
 \begin{equation}
 S_+ =2\pi \int r_+ dr_+ ,
 \end{equation}
 which can be integrated to give
 \begin{equation}
 S_+ =\pi r_+^2.
 \end{equation}
 
 \begin{figure}[hbtp]
\begin{center}	
\includegraphics[width=0.48\textwidth]{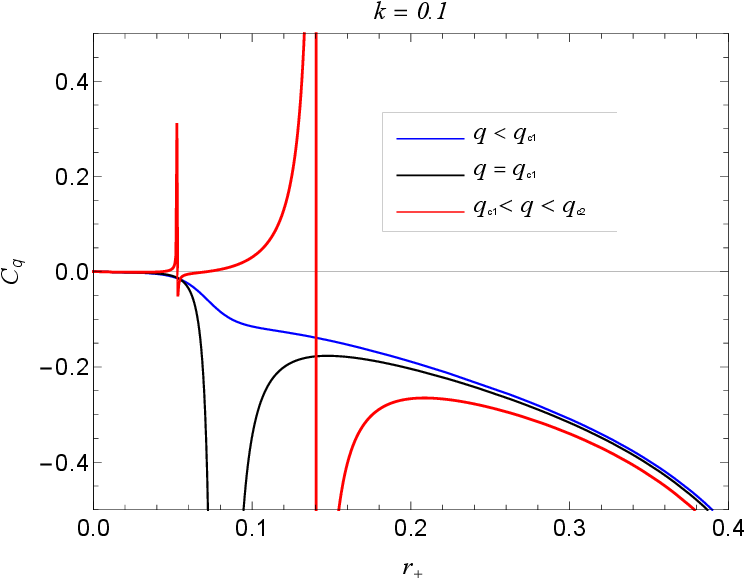}
\includegraphics[width=0.48\textwidth]{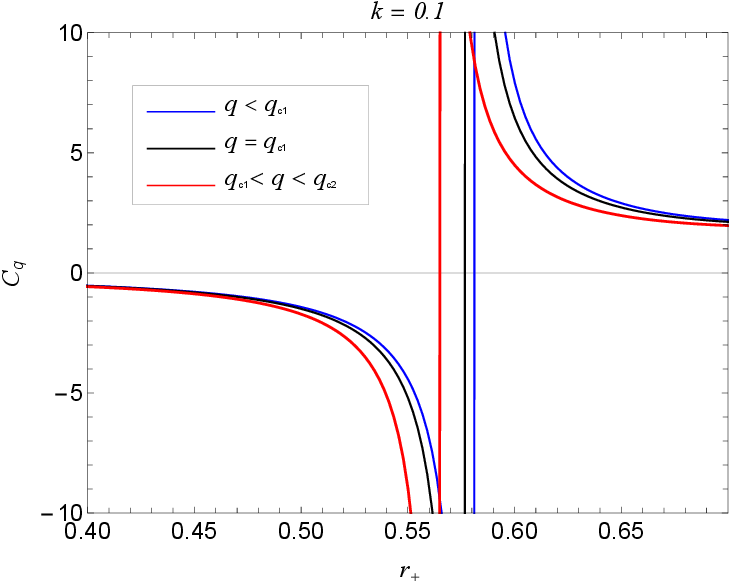}\\
\includegraphics[width=0.48\textwidth]{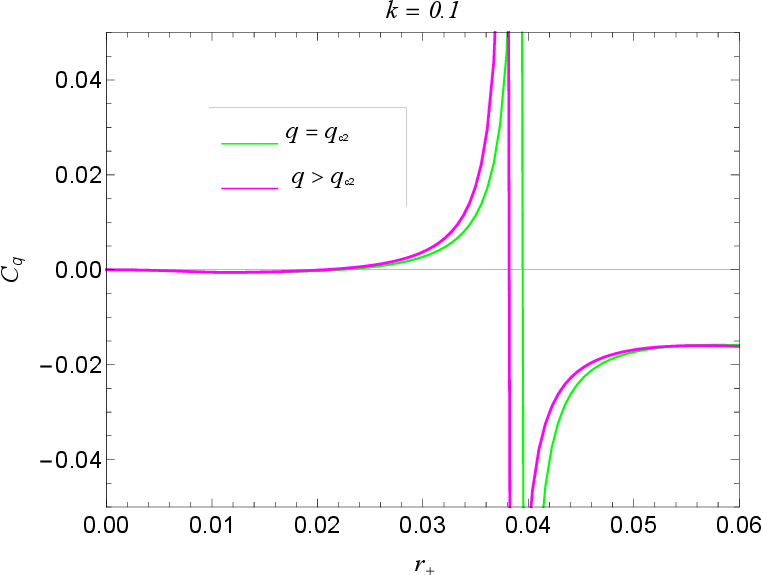}
\includegraphics[width=0.48\textwidth]{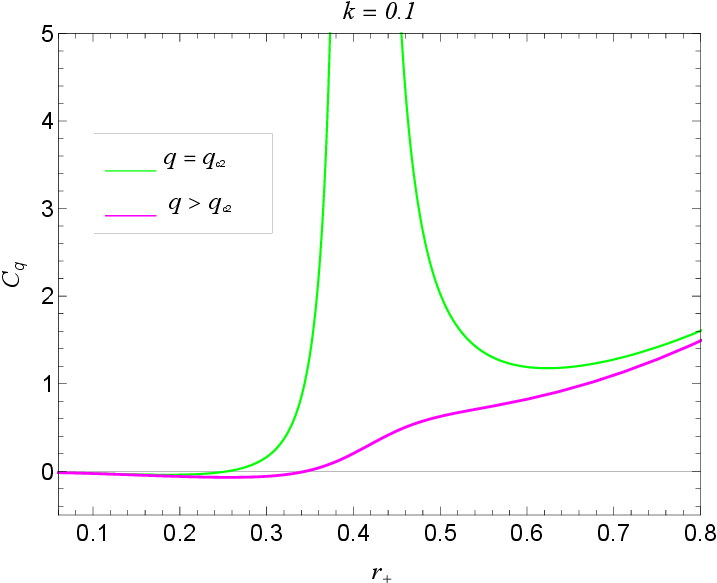}
\caption{
 Specific heat $C_q$ vs horizon radius $r_+$ for different $q$ with $k = 0.1$, $\alpha_3 = 0.1$, $\alpha_4 = 0.2$. The positive (negative) $C_q$ indicates local stability (instability). Divergences mark phase transitions.
}
\label{plotcq}	
\end{center}   
\end{figure}
 Very interestingly, this formula is identical to that found in Ref. \cite{Ghosh:2015cva} with or without a linear electromagnetic field. This formula can be rewritten to be the famous one, in which $S_+ = \frac{A_+}{4}$ with $A_+ \equiv 4\pi r_+^2$ being the area of the horizon spherical surface. This implies that the presence of the massive graviton does not affect the entropy formula of the black hole. It just modifies the value of the event horizon radius $r_+$. We next turn to the local stability or thermal stability. The reason to consider local stability is that even when a black
hole configuration is globally stable, it can be locally unstable or vice versa. One can investigate the thermodynamic stability from the specific heat of the system \cite{Kubiznak:2014zwa,Myrzakulov:2023rkr,Ditta:2024jrv}. The specific heat at constant charge $C_q$  defined as (see Fig. \ref{plotcq} for the behavior of the $C_q$ versus $r_+$)
\begin{align}  \label{cq} 
 C_q \equiv T \Big(\frac{\partial S}{\partial T}\Big)_{q}=
  \frac{2 \pi r_+^3 \left\{  q^2\exp\left[\frac{-k}{r_+} \right] -  r_+^2 \left( 1 +\zeta + 2 \gamma r_+  + 8 \pi P r_+^2  \right) \right\}}{ q^2\exp\left[\frac{-k}{r_+} \right]\left(k-3r_+\right) +   r_+^3 \left( 1 + \zeta -8\pi P r_+^2  \right)}.
\end{align}
 %%%%%%%%%%%%

It is a well-known fact that the positivity of the specific heat, along with positive temperature, signifies the thermal stability of the system. When a black hole is unstable, it can undergo a phase transition to reach a stable state. We aim to find the phase transition point by identifying the roots and divergences of $C_q$. The roots of the heat capacity, which correspond to zero temperature, mark the thermal transition between the unphysical $(T_+<0)$ and physical $(T_+>0)$ states of the black hole. Also, according to thermal physics, a divergence in heat capacity signals the existence of a phase transition. Because it is difficult to find the roots and divergences of the heat capacity analytically, we use the numerical results shown in Fig. \ref{plotcq} to study this. Fig. \ref{plotcq} shows how $C_q$ changes with $r_+$ for different parameter sets. The stability clearly depends on the metric parameters. The behaviour of the specific heat shown in Fig. \ref{plotcq} leads us to the following conclusions:
\begin{itemize}
    \item When $q< q_{c1}$ (blue curves) or $q>q_{c2}$ (magenta curves), there are two phases, one unstable ($C_q<0$) and one stable ($C_q>0$), and there exists a phase transition between the unstable and stable configurations at the global minima of the temperature or the point of divergence of $C_q$. Importantly, the specific curves for $q>q_{c2}$ exhibit two extra phases (one with $C_q>0$ and the other with $C_q<0$) which exist on either side of the divergence point of $C_q$, but these phases have negative temperature; therefore, these phases are unphysical. 
    \item For $q_{c1}<q<q_{c2}$ (red curves), although the Fig. shows six different phases out of which three phase have $C_q>0$ whereas the other three have $C_q>0$, but the two phases, one with positive and the other with negative specific heat on either side of the first divergence point of $C_q$ are unphysical as the temperature for those phases is negative. Hence, we can confirm that for this range of charge, we have two stable and as many unstable phases. 
    \item For $q=q_{c1}$ (black curves) or $q=q_{c2}$ (green curves), there exist three different phases such that when $q=q_{c1}$, there are two unstable phases ($C_q<0$) and one stable phase ($C_q>0$) while for $q=q_{c2}$ the specific heat is positive for two phases (stable) and negative for one phase (unstable). For $q=q_c2$, the specific heat exhibits a cusp at the second critical point ($r_{c2},T_{c2}$), and a second-order phase transition takes place between two stable phases. Also, there exist two extra unphysical phases ($T<0$) when $q=q_{c2}$.   
\end{itemize}

\subsection{Global Stability and Phase Structure}\label{3B}
Next, we examine how Gibbs free energy behaves in our black holes to understand the system’s phase structure. This energy helps identify which state is most stable at equilibrium. The state with the lowest Gibbs free energy is the most stable overall. The Gibbs free energy of the black hole is calculated, according to Ref. \cite{Kumar:2018vsm}, as follows
\begin{equation}
    G_+ = M_+ - T_+S_+,
\end{equation}
which comes out to be
\begin{equation}
G_+ = \frac{1}{12} \left\{ \frac{3q^2  \exp\left[-\frac{k}{r_+} \right]  \left(k - 2 r_+ \right)}{k r_+} + r_+ \left(3 + \Lambda r_+^2 + 3 \zeta\right) \right\},
\end{equation}
which will reduce to the Gibbs free energy of Schwarzschild-AdS black holes  in the setting $q=0$ \cite{Cai:2001dz},
\begin{equation}\label{29}
G_+ = \frac{1}{12} \, r_+ \left(3 + \Lambda r_+^2 + 3 \zeta\right).
\end{equation}
    \begin{figure}[hbtp]
\begin{center}	
\includegraphics[width=0.32\textwidth]{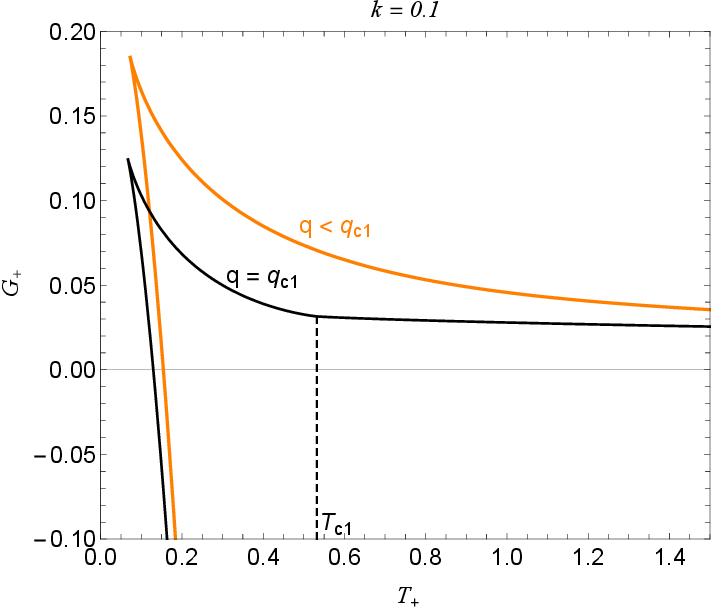}
\includegraphics[width=0.32\textwidth]{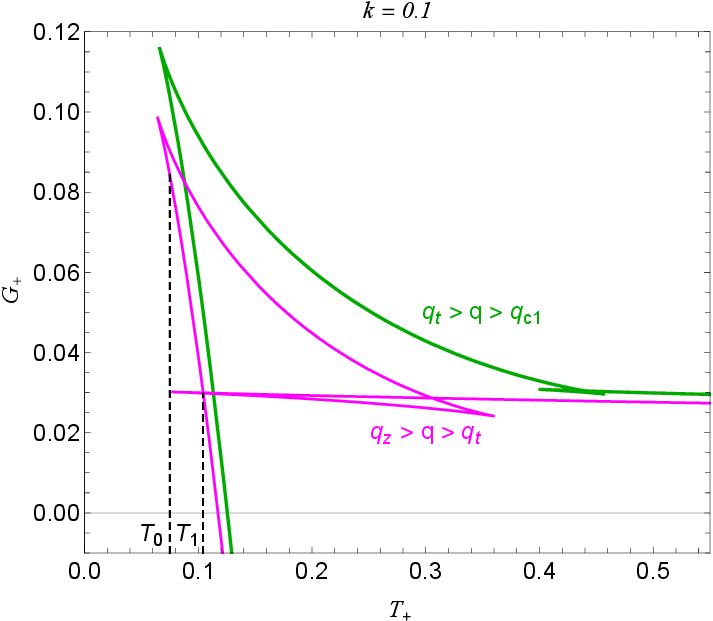}
\includegraphics[width=0.32\textwidth]{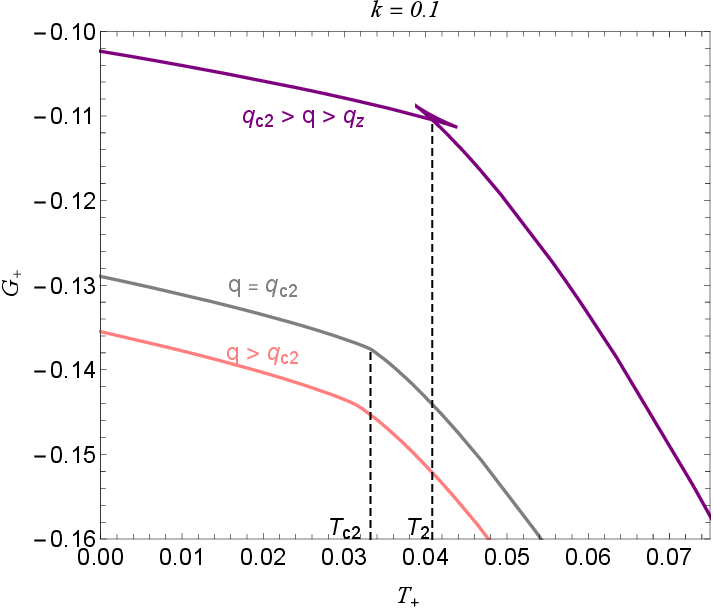}
\caption{Gibbs free energy $G_+$ vs temperature $T_+$ for different $q$ with $k = 0.1$, $\alpha_3 = 0.1$, $\alpha_4 = 0.2$. Swallowtail structures indicate first-order transitions, cusps indicate second-order transitions.}
\label{plotgt}
\end{center}   
\end{figure}
  \begin{figure}[hbtp]
\begin{center}	
\includegraphics[width=0.5\textwidth]{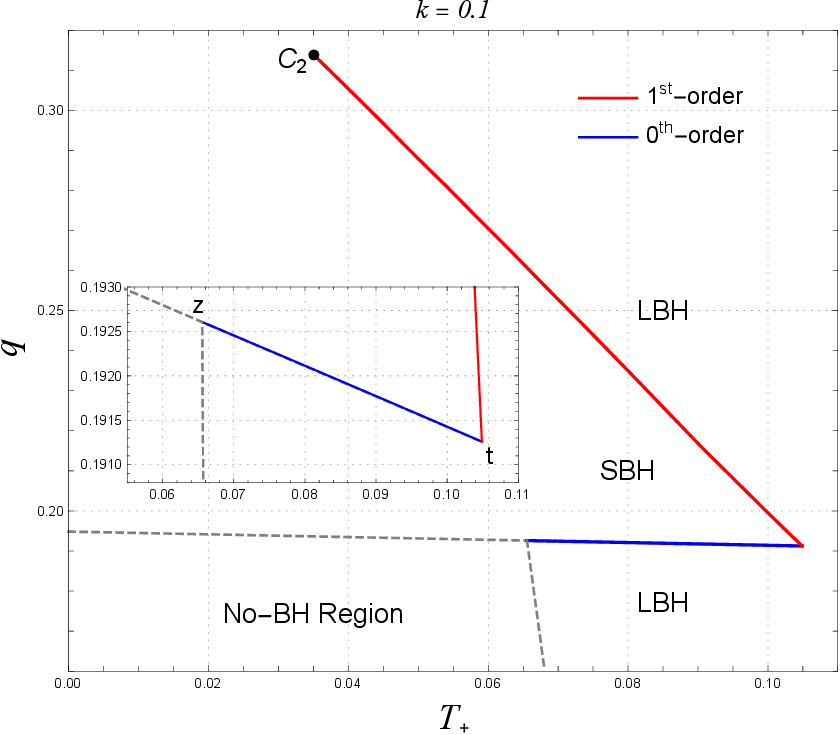}\\
\caption{Phase diagram showing coexistence curves for first-order (red) and zeroth-order (blue) phase transitions. Points $z$ and $t$ bound the reentrant phase transition region. The area between the dotted lines denotes no black hole solutions.}
\label{plotgt2}
\end{center}   
\end{figure}
%   \begin{figure}[hbtp]
% \begin{center}	
% \includegraphics[scale=0.5]{gtr.pdf}\\
% \caption{Schematic representation of the small/large black hole phase transition analogous to liquid-gas systems. The diagram shows transition temperatures $T_1$ and $T_2$, with the first-order phase transition occurring at $T_0$ where the Gibbs free energies of small and large black hole phases are equal.}
% \label{plotgt2}
% \end{center}   
% \end{figure}

Since the global stability of the system is determined
by the Gibbs free energy, its global minimum is estimated to be the preferred state of the black hole \cite{Kubiznak:2014zwa,Kubiznak:2016qmn,Yasir:2023ysw,Ditta:2023luy}.
In Fig.~\ref {plotgt}, the behaviour of the Gibbs free energy $G_+$ versus temperature $T_+$ of the charged AdS black hole in massive gravity has been presented. One can observe from Fig.~\ref{plotgt} that the Gibbs free energy exhibits different characteristics for different values of the charge. We have shown seven different $G_+-T_+$ curves depending upon the different ranges of the charge, which help us to understand the phase structure of our system as follows:
\begin{itemize}
    \item $\mathbf{q<q_{c1}}$ (orange curves): For this range of $q$, there exist two branches; one thermally stable and other unstable. The Gibbs free energy of the stable branch is always less than that of the unstable branch; hence, we can conclude that the system will remain in the stable branch and no phase transition exists for this case.
    \item $\mathbf{q=q_{c1}}$ (black curves): There are two unstable branches and one stable branch for this case. The $G_+-T_+$ curves show a cusp at $T_{c1}$ (first critical point), which suggests a second-order phase transition between the two unstable branches, but the stable branch always has a smaller Gibbs free energy than that of the unstable branches. That means the system's preferred state is the stable branch, and hence the phase transition between the two unstable branches is unphysical. Therefore, for $q=q_{c1}$, our system exhibits no phase transition.
     \item $\mathbf{q_{c1}<q<q_{t}}$ (green curves): There are two stable and two unstable branches, but the Gibbs free energy of the large black holes is minimum, which means the system will remain in this state, and hence there exists no phase transition.
     \item $\mathbf{q_{t}<q<q_{z}}$ (pink curves): A notable feature observed in this range of the charge parameter is the reentrant phase transition, where two successive phase transitions occur as a single thermodynamic variable is varied, ultimately returning the system to a macroscopic state similar to the initial one \cite{Altamirano:2013ane,Frassino:2023wpc,Gunasekaran:2012dq,Fernando:2016qhq,Chaloshtary:2019qvv,Zou:2016sab}. When the temperature satisfies $T_+>T_1$, the system exists in a stable large black hole (LBH) phase. As the temperature decreases to $T_+=T_1$, the LBH undergoes a first-order phase transition to a stable small black hole (SBH). Upon further lowering the temperature, a discontinuous jump in the Gibbs free energy appears at $T_+=T_0$, indicating a zeroth-order phase transition that transforms the stable SBH back into the LBH phase. This zeroth-order transition begins at the point$z$ ($q_z$, $T_z$) and ends at the point $t$ ($q_t$, $T_t$) as illustrated in Fig. \ref{plotgt2}. For the parameter choice $k=0.1$, $\alpha_3=0.1$, and $\alpha_4=0.2$ the coordinates of the points $z$ and $t$ are found to be ($0.1926, 0.06555$) and ($0.19126,0.104869$), respectively.
     \item $\mathbf{q_{z}<q<q_{c2}}$: stable SBH undergoes a Van der Waals-like first-order phase transition to stable LBH at $T_+=T_2$ that is characterized by the swallowtail structure of $G_+-T_+$ curves. 
     \item $\mathbf{q=q_{c2}}$: The stable SBH undergoes a second-order phase transition to stable LBH at $T_+=T_{c2}$ that is characterized by a cusp of $G_+-T_+$ curves.
     \item $\mathbf{q>q_{c2}}$: The system remains in stable LBH phase, and hence no phase transition occurs.
\end{itemize}

We have depicted the complete phase structure of charged AdS black holes in massive gravity on $G_+-T_+$ planes in Fig. \ref{plotgt2}. The phase structure analysis reveals three types of phase transitions: zeroth-order, first-order, and second-order. The red curve in Fig. \ref{plotgt2} represents the coexistence curve of a first-order phase transition starting from point $t$ and ending at the second critical point $C_2$ such that above the curve we have LBH, whereas the SBH phase is below the curve. Between the points $t$ and $z$, there exists a zeroth-order phase transition accompanying the first-order phase transition. This phenomenon is known as the reentrant phase transition, specifically the LBH/SBH/LBH type. The blue curve starting from $z$ and ending at $t$ signifies the coexistence curve of LBH and SBH for a zeroth-order phase transition, so that the region below the curve belongs to the LBH phase, while the region above the curve denotes the LBH phase. The No-BH Region denotes the parameter space where no black hole exists.

%%%%%%%%%%%%%%%%%%%%%

%%%%%%%%%%%%%%%%
\section{Conclusions} \label{final}
  %%%%%%%%%%%%%%%%%%%%%%%%%%%%%%%%%%
We present a novel class of exact charged AdS black hole solutions in 4D dRGT massive gravity, minimally coupled to a  NED model described by an exponential Lagrangian. Interestingly, the dRGT massive gravity provides a ghost-free framework with a finite graviton mass $m_g$, and the graviton potentials actually mimic a cosmological constant $\Lambda$, alongside two additional constants \( \gamma \) and \( \zeta \). These parameters collectively modify the spacetime geometry of the black holes. The exponential NED model introduces a magnetic monopole charge $q$ and a nonlinearity parameter \( k \). The black hole solution reduces to known cases in appropriate limits (e.g., the Schwarzschild solution when $m_g \to 0$ and $ q \to 0$). Interestingly, and in contrast to several other NED black hole models that produce regular black holes, our resulting geometry in this analysis remains singular at the origin \( r = 0 \),  emphasising a key distinctive feature of this solution. The combined effects of massive gravity and NED lead to a black hole solution that is distinct from previous results in both pure dRGT and NED contexts.
The main characteristic of the obtained black holes is based on the derived $F(r)$ function shown in Eq. \eqref{sol-of-F-new} for the choice of exponential electrodynamics shown in Eq. \eqref{exponential-electrodynamics}, which has been considered in a number of previous papers \cite{Nam:2018ltb,Ghosh:2018bxg,Ghosh:2020ijh}. As a result, the obtained black holes are different from those found in Ref. \cite{Ghosh:2015cva} as well as Ref. \cite{Nam:2018ltb} in some limits. More importantly, they are not regular black holes, in contrast to those found in NED models \cite{AyonBeato:1998ub,Ayon-Beato:1999kuh,Bronnikov:2000vy,AyonBeato:2000zs,Fernando:2016ksb,Ghosh:2018bxg,Ali:2018boy,Kumar:2018vsm,Hyun:2019gfz,Nomura:2020tpc,Ghosh:2020ijh,Hendi:2020knv,Guo:2021zxl,Rehan:2024dsg,Sudhanshu:2024wqb}. 

 The fundamental quantities—mass \( M_+ \), Hawking temperature \( T_+ \), and entropy \( S_+ \)—were also analysed. The entropy steadfastly adheres to the Bekenstein-Hawking area law, \( S_+ = \pi r_+^2 \), indicating that the massive graviton does not modify the fundamental area law. The entropy remains unaffected by the graviton mass, depending only on the horizon area. In turn, we discussed the thermodynamic stability of the system through the numerical analysis of the specific heat at constant charge ($C_q$) shown in Fig. \ref{plotcq}. We provided a detailed analysis of $C_q$, in which we found that there exist many stable and unstable phases depending on the value of $q$. We observed the divergence of $C_q$, signifying the phase transitions between stable and unstable phases.

The behavior of the Gibbs free energy $G_+$ provides important insight into the thermodynamic phase structure of the system. Our analysis shows that, depending on the value of the charge parameter $q$, the black hole system can exhibit three different types of phase transitions: zeroth-order, first-order, and second-order transitions. When the charge is either larger than $q_{c2}$ (red curve in Fig.~\ref{plotgt}) or smaller than $q_t$ (orange, black, and green curves in Fig.~\ref{plotgt}), the system possesses only a single thermodynamically stable phase, and therefore no phase transition occurs. At the critical value $q=q_{c2}$, the small black hole (SBH) undergoes a second-order phase transition to a large black hole (LBH) at the second critical temperature $T_+=T_{c2}$. For charge values in the interval $q\in(q_z,q_{c2})$, the system experiences a van der Waals--like first-order phase transition from a stable SBH to a stable LBH. This transition begins at point $t$ and terminates at the second critical point $C_2$, as illustrated by the red curve in Fig.~\ref{plotgt2}. Moreover, within the charge range $q\in(q_t,q_z)$, the system exhibits a more intricate behavior. In this region, the black hole undergoes a zeroth-order phase transition between the LBH and SBH phases, characterized by a finite discontinuity in the Gibbs free energy. This transition occurs together with a first-order SBH/LBH phase transition, resulting in a reentrant phase transition of the form LBH/SBH/LBH, where the initial and final macroscopic states are identical. The reentrant transition begins at point $z$ and ends at point $t$. The coexistence line corresponding to the zeroth-order phase transition is represented by the blue curve in Fig.~\ref{plotgt2}.

By demonstrating how a NED and a ghost-free massive gravity theory can interact to produce novel physical phenomena, this investigation enhances our understanding of black hole thermodynamics. The findings provide a rich field for further investigation and support a compelling thermodynamic comparison between black holes and more well-known systems, such as the van der Waals fluid.   The study reinforces the deep analogy between black hole thermodynamics and conventional thermodynamic systems, particularly through the van der Waals-like and reentrant phase structure. The results are relevant in the context of the gauge/gravity duality, as AdS black holes are dual to finite-temperature conformal field theories.
The physical importance of our model is multifaceted. Observational constraints on the graviton mass $m_g$ would directly bound the values of the derived parameters, $\Lambda$, $\gamma$, and $\zeta$, thereby restricting the admissible thermodynamic phase space of these black holes. Potential astrophysical signatures may arise from deviations in the shadow cast by these black holes or their quasi-normal mode, driven by the massive gravity corrections.

The present work offers several avenues for future research that would further elucidate the properties of black holes in dRGT massive gravity coupled to NED. A natural extension would be to investigate the dynamical stability of these solutions through linear perturbation analysis, examining both scalar and gravitational perturbations to investigate their stability in the parameter space. Additionally, an interesting avenue is to examine the holographic implications of these black holes within the AdS/CFT correspondence, specifically the effects of the huge gravity parameters on the transport characteristics and thermalisation dynamics of the dual field theory. Furthermore, investigating the rotating or higher-dimensional generalisations, as well as holographic applications such as conductivity or entanglement entropy. The influence of other NED models or other graviton potentials.

In summary, this paper offers a consistent and rich thermodynamic description of a new family of charged AdS black holes in dRGT massive gravity with exponential NED, demonstrating critical behaviour and phase transitions that closely mirror those of classical thermodynamic systems.

%%%%%%%%%%%%%%%%%%%%%%
\begin{acknowledgments}
This work was supported by the SERB-DST ASEAN project No. CRD/2018/000042. T.Q.D. is funded by the Vietnam National Foundation for Science and Technology Development (NAFOSTED) under grant number 103.01-2023.50. T.Q.D. would like to thank the Centre for Theoretical Physics of Jamia Millia Islamia very much for its warm hospitality.  
\end{acknowledgments}
%%%%%%%%%%%%%%%%%%%%%%%%%%%%%%%%%%%%%%%%%%

\bibliographystyle{apsrev4-1}
\bibliography{MG}
\end{document}